\documentclass[10pt, twocolumn]{article}

\usepackage{times}
\usepackage[scaled=.90]{helvet}
\usepackage{courier}
\usepackage{graphicx}
\usepackage[tight]{subfigure}
\usepackage{xspace}
\usepackage{url}
\usepackage{cite}

\newcommand{\eg}{\emph{e.g.,}\xspace}
\newcommand{\ie}{\emph{i.e.,}\xspace}
\newcommand{\etal}{\emph{et al.}\xspace}
\newcommand{\paratitle}[1]{\vspace{1.5ex}\noindent \textbf{#1}}
\newcommand{\sdf}{S\textsuperscript{3}D\xspace}

\title{Semi-Supervised Spam Detection in Twitter Stream}

\author{Surendra Sedhai \hspace{1in} Aixin Sun \\
School of Computer Science and Engineering\\
Nanyang Technological University, Singapore}
\date{\textsf{surendra001@e.ntu.edu.sg}\hspace{1in} \textsf{axsun@ntu.edu.sg}
 }
 
\begin{document}

\twocolumn[
  \begin{@twocolumnfalse}
\maketitle

\begin{abstract}
Most existing techniques for spam detection on Twitter aim to identify and block users who post spam tweets. In this paper, we propose a Semi-Supervised Spam Detection (S\textsuperscript{3}D) framework for spam detection at tweet-level. The proposed framework consists of two main modules: \textit{spam detection} module operating in real-time mode, and \textit{model update} module operating in batch mode. The spam detection module consists of four light-weight detectors: (i) blacklisted domain detector to label tweets containing blacklisted URLs, (ii) near-duplicate detector to label tweets that are near-duplicates of confidently pre-labeled tweets, (iii)  reliable ham detector to label tweets that are posted by trusted users and that do not contain spammy words, and (iv) multi-classifier based detector labels the remaining tweets. The information required by the detection module are updated in batch mode based on the tweets that are labeled in the previous time window. Experiments on a large scale dataset  show  that the framework adaptively learns  patterns of new spam activities and maintain good accuracy for spam detection in a tweet stream.
\end{abstract}

\vspace{5ex}
  \end{@twocolumnfalse}
]

\section{Introduction}\label{sec:intro}

Micro-blogging services have attracted the attention of not only legitimate users but also spammers.  It is reported that 0.13\%  of messages advertised on Twitter are clicked, which is two orders of magnitude higher than that of email spam~\cite{Grier:2010}. High click-rate and effective message propagation make Twitter an attractive platform for spammers. Increasing spamming activities have adversely affect user experience as well as many tasks such as user behaviour analysis and recommendation.

Most of the existing studies on Twitter spam focus on account blocking, which is to identify and block spam users, or spammers. Hu \etal utilized social graph and  the tweets of a user and formulated spammer detection as an optimization problem~\cite{Hu:2013}. Similarly, information extracted from user's tweets, demographics,  shared URLs, and social connection are utilized as features in standard machine learning algorithms to detect spam users~\cite{Lee:2010}.  However, account blocking approach is less effective for spammers who may act as legitimate users by posting non-spam content regularly. Blocking spammers may even hurt a legitimate user who happens to grant permission to a third-party application that posts spammy tweets under her username. This legitimate account may be blocked because of such spam tweets. Furthermore, spammers change their tweet content and strategies  to make their tweets  and activities look like legitimate~\cite{Benevenuto:2010}. While identifying and blocking spammer accounts remain a crucial and challenging task, tweet-level spam detection is essential to fight against spamming at a more fine-grained level, and helps to timely detect spam tweets instead of waiting for users to be detected as spammers. Similarly, Chen \etal suggested that training dataset should be continuously updated in order to deal with changing  distribution of features in tweet stream~\cite{Chen:2015}.

In this paper, we propose a semi-supervised framework for spam tweet detection. The proposed framework mainly consists of two main modules: (i) four light-weight detectors in the \textit{spam tweet detection module} for detecting spam tweets in real-time, and (ii) \textit{updating module} to periodically update the detection models based on the confidently labeled tweets from the previous time window. The detectors are designed based on our observations made from a collection of 14 million tweets, and the detectors are computationally effective, suitable for real-time detection. More importantly, our detectors utilize classification techniques at two levels, tweet-level and cluster-level. Here, a cluster is a group of tweets with similar characteristics. With this flexible design, any features that may be effective in spam detection can be easily incorporated into the detection framework.  The framework starts with a small set of labeled samples, and update the detection models in a semi-supervised manner by  utilizing the confidently labeled tweets from the previous time window. This semi-supervised approach helps to learn new spamming activities, making the framework more robust in identifying spam tweets.

\section{Related Work}
\label{sec:related}

Spam is a serious problem on almost all online media and spam detection has been studied for many years. Spammers may use different techniques on different platforms so spam detection technique developed for one platform may not be directly applicable on other platforms. Thomas \etal~\cite{Thomas:2011} reported that spam targeting email is significantly different from spam targeting Twitter. In Twitter, there are different types of spamming activities such as link farming~\cite{Ghosh:2012}, spamming trending topics~\cite{Benevenuto:2010},  phishing~\cite{Chhabra:2011}, aggressive posting using social bot~\cite{Chu:2010}, to name a few. These different activities pollute timeline of users as well as Twitter search results.

Many  social spam detection studies focus on the identification of spam accounts. Lee \etal~\cite{Lee:2010} analyzed and used features derived from user  demographics, follower/following social graph, tweet content and temporal aspect of user behavior to identify content polluters.  Hu \etal exploited social graph and  tweets of a user to detect spam detection on Twitter. They formulated spammer detection task as an optimization problem~\cite{Hu:2013}. Online learning has been utilized to tackle the fast evolving nature of spammer~\cite{Hu:2014b}. They have utilized both content and network information and incrementally update their spam detection model for effective social spam detection. Tan \etal~\cite{Tan:2013} proposed an unsupervised spam detection system that exploits legitimate users in the social network.  Their analysis shows the volatility of spamming patterns in social network. They have utilized non-spam patterns of legitimate users based on social graph and user-link graph to detect spam pattern. Gao \etal~\cite{Gao:2010} identified social spam by clustering posts based on text and URL similarities and detecting large clusters with bursty posting patterns.

Removing  spam users cannot filter every spam message as spammer may create another account and restart spamming activity.  This calls for tweet-level spam detection. Inspired from content-based techniques for emails, \cite{Igor:2013} utilized standard classifiers to detect spam tweets. Language modeling approach has been used to  compute the divergence of trending topic, suspicious message, and title of the page linked in the tweet~\cite{Martinez:2013}. Similarly, Castillo \etal~\cite{Castillo:2011} analyzed the credibility of tweets on trending topics based on users' tweeting and retweeting behaviors, tweet content and link present in the tweets. As spammers keep on evolving over time, hence semi-supervised approach is suitable for tracking such changing spamming activities.  Semi-supervised spam detection approach has been utilized to identify spam on voice-over-IP  call~\cite{Wu:2009}.  Similarly, a semi-supervised approach is reported to have better performance than supervised approach for malware detection task~\cite{Santos:2011}. However, to the best of our knowledge semi-supervised approach has not be utilized to detect spam tweets. Our proposed method is capable of continuously updating itself by using semi-supervised approach.


\begin{figure}[t]
 \centering
  \includegraphics[width=3.45in]{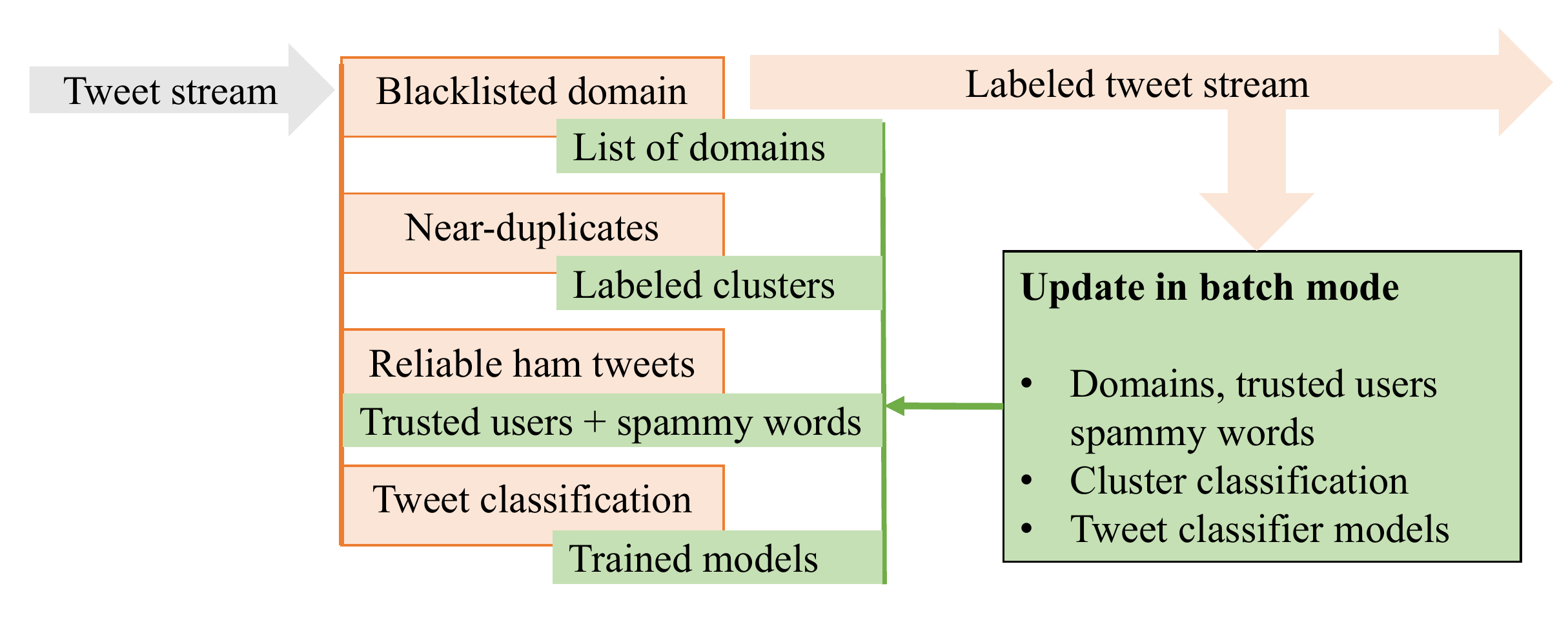}
  \caption{System Overview of the \sdf framework.}
  \label{fig:systemOverview}
\end{figure}

\section{Semi-supervised Spam Detection}
\label{sec:systemDescription}

The proposed \sdf contains two main modules as shown in Figure~\ref{fig:systemOverview}. Assuming that we have all the information (\eg a blacklist of spamming domains, and trained classification models), the tweets are labeled as spam and non-spam (also known as  ``ham'') tweets using the four detectors in real-time. The required information is updated periodically based on the confidently labeled tweets from the previous time window, in a semi-supervised manner. Next, we detail the main modules.

\subsection{Spam Tweet Detection in Real-Time}
For efficiency reason, the tweets are labeled by four light-weight detectors from four perspectives, in an order of the easiest to hardest in terms of difficulty in detection. Once a label is assigned by one of the detectors, the tweet need not pass to the next detector.

\paratitle{Blacklisted Domain Detector.} Spammers promote their services/products by posting links in their tweets~\cite{Gao:2010}. An effective way of spam detection is to detect tweets containing links from blacklisted domains. The list of blacklisted domains is to be updated at the end of each time window utilizing confidently labeled tweets, during batch update.

\paratitle{Near-Duplicate Detector.} Tweets that are near-duplicates of pre-labeled spam/ham tweets are assigned the same labels accordingly. The near-duplicate tweets are detected by using the \textit{MinHash} algorithm~\cite{Broder:1997}, which has shown effectiveness for labeling spam tweets~\cite{Sedhai:2015}. More specifically, a signature is computed for each tweet by concatenating the three minimum hash values computed from the tweet's \textit{uni-gram}, \textit{bi-gram}, and \textit{tri-gram} representations, respectively. If two or more tweets have the same signature, then the tweets are considered near-duplicates. If a cluster of near-duplicate tweets hashed to the same signature has been labeled as spam or ham tweets, the new tweet having the same signature receives the same label.

\paratitle{Reliable Ham Tweet Detector.} Tweets posted by legitimate users can be considered as ham tweets, however; spammers may pretend as legitimate users and after gaining acceptance from other users they post spam tweets~\cite{Yang:2011}. Hence, we consider a tweet to be a \textit{reliable ham tweet} if it satisfies two conditions: (i) the tweet does not contain any spammy words, and (ii) the tweet is posted by a trusted user.

Spammy words are that words whose probability of occurrence is larger in spam than in ham tweets. For example, word \textit{followme} is likely to appear in spam tweets but the word may appear in ham tweet as well. Specifically, let $p_s(w)$ be the probability of word $w$ appearing in the spam tweets, and $p_h(w)$ be the probability of $w$ in ham tweets. Then $w$ is a spammy word if $p_s(w) > p_h(w)$. In our implementation, words that are shorter than 3 characters in length are ignored. 

A trusted user is a user who has never posted any spam tweet and has posted as least 5 confident ham tweets. A tweet is a confident ham tweet if the tweet does not contain any spammy words and is predicted to be ham by all the component classifiers in the tweet classification detector. The clusters of near-duplicate tweets will also be predicted as ``clusters of spam'' or ``clusters of ham'' by multiple classifiers. The tweets in a cluster which is predicted to be ham cluster by all classifiers are also considered as confident ham tweets.

During the batch update, the list of trusted users and the list of spammy words are updated. 

\paratitle{Multi-classifier based Detector.} Tweets that are not labeled in any of the previous steps are processed and labeled in this step. Here, we develop a spam detector by using three efficient classifiers, namely Na\"{\i}ve Bayes (NB), Logistic Regression (LR), and Random Forest (RF). The three classifiers use different classification techniques, \ie generative, discriminative, and decision tree-based classification models. A full spectrum of features is extracted to represent each tweet. Listed in Table~\ref{tab:features} in the column titled ``Features for tweet representation'', the features include hashtag-based features, content-based features, user-based features, and domain-based features. Most features are self-explanatory and we only elaborate two features, categorical hashtag, and top domains. Categorical words are the words used in one of the top-level categories in Yahoo! hierarchy, or words used to categorize content in four Web sites: BBC, CNN, NYTimes and Reddit. There are 75 categories including sports, technology, business, movie, jobs etc. The binary feature is 1 if the hashtag is one of the categorical words. The domain feature is based on the domain of the URLs contained in tweets. Domain ranking is from alexa.com. A tweet is labeled as spam if at least two of the three classifiers predict the tweet to be spam; otherwise the tweet is labeled as ham.

\subsection{Model Update in Batch Mode}
\label{subsec:batchUpdate}

The time window for the update is set to be \textit{one day} in our experiments. The key desideratum is to identify the confidently labeled data of the previous time window.

\paratitle{Confidently Labeled Tweets.}
Tweets that are labeled by the first three detectors (\ie  blacklisted domain, near-duplicate, and reliable ham tweet) are considered as confidently labeled tweets. For the classifier based detector, recall that we use three classifiers each is based on a different classification technique. Tweets that are labeled as spam by all the three classifiers are considered as confidently labeled spam tweets. Similarly, tweets that do not contain any spammy words and are labeled as ham by all the three classifiers are confidently labeled ham tweets. Excluding ham tweets containing spammy words (\eg \textit{followme}) helps to prevent the deviation of classifier from a burst of spammy words in ham tweets.

The identified confidently labeled spam tweets are utilized to update \textit{blacklist domains}, and confidently labeled ham tweets are utilized to identify \textit{trusted users}.

\paratitle{Near-duplicate Cluster Labeling.} Recall that the near-duplicate detector computes a signature for each tweet to check if the tweet is a near-duplicate  of a labeled cluster. If the signature of a tweet  does not match any pre-labeled cluster, then the tweet is passed to the next level detectors.

After each time window, all the tweets that do not match pre-labeled clusters but having the same signature are grouped into a new cluster; \ie each cluster is a collection of near-duplicate tweets. Next, we label the clusters each containing at least 10 tweets and if the labels are of high confidence, then the signatures of these newly labeled confident clusters will be used by the near-duplicate detector in the next time window.  Recall that all the tweets have been labeled as spam and ham tweets (see Figure~\ref{fig:systemOverview}), an easy approach to label these clusters is to perform a majority voting. Specifically, if there are more spams in a cluster than ham tweets, then the cluster is labeled as a spam cluster. However, the majority voting approach solely relies on the predicting power of the detectors and may not capture the new spamming patterns in the most recent time window. Moreover, because tweets in a cluster are near-duplicates, their labels assigned by the detectors are mostly the same. For this reason, we also employ a feature-based classifier.

Each cluster is represented with hashtag-based features, content-based features, user-based features,  and domain-based features, as listed in Table~\ref{tab:features}, the third column. Many of the features used here are adopted from existing studies~\cite{Castillo:2011,Ferrara:2014}. Different from tweet classification (features listed in the second column), the cluster-level  features represent the collective information obtained from all the tweets in the cluster. The clusters represented in feature-space are classified  using a logistic regression classifier.

We consider a cluster to be a confidently labeled cluster if the labels predicted by the feature-based cluster classifier and the majority voting of the tweet labels are the same.

\paratitle{Update Detector Models.} After finding the confidently labeled tweets and clusters,  the models used by the detectors are updated accordingly including blacklisted domains, labeled clusters, trusted users and tweet classification models. Blacklisted domains are updated by including domains having at least 5 tweets in the last time window and  at least 90\% of the tweets are confidently labeled as spam tweets. A user having at least 5 tweets and all tweets are confidently labeled ham tweets is considered as a trusted user. The  classification  models of the three classifiers are retrained  by including the newly labeled confident tweets of the last time window.

By updating the detection models in batch mode, the proposed semi-supervised spam detection  framework is capable of capturing new vocabulary and new spamming behaviours, which makes the framework robust and adaptive to deal with the dynamic nature of spamming activities.

\subsection{Computational Efficiency}

All the four spam detectors are computational effective, hence the proposed framework is capable of  labeling tweet stream in real-time.
We conduct experiments on a desktop PC  with octa-core Intel processor of 3.70GHz and 16 GB RAM. In our experiments, all the detectors are carried out on a single-core of the processor, except random forest classifier which utilizes all the cores of the processor. Empirically, we found that on average it takes 0.495 ms to label a tweet where more than 50\% of the time is used for feature extraction. Note that, our code is not optimized for real-time setting, and efficiency can be further improved by parallelising the detectors.

\section{Experiment and Discussion}
\label{sec:experiment}

We used 15 days of data from HSpam14 dataset~\cite{Sedhai:2015} in our experiments. HSpam14 contains 14 million tweets, collected by using the trending topics on Hashtags.org for two months, May and June 2013.  In this research, we use 15 days of tweets (May 17 - 31, 2013) where each day has more than 35 thousand tweets. The time window for batch mode update is set to be a day. Almost all tweets in HSpam14 are labeled to be spam and ham and the remaining small portion are labeled as unknown for not being able to determine their labels even with manual inspection. Note that, more than 80\% of tweets in HSpam14 are labeled  automatically and the manually labeled tweets are biased to spams.

We simulate a tweet stream in our experiments. On the first day, the detectors in \sdf are trained by using the manually labeled tweets and the reliable ham tweets in the HSpam14 dataset (the released HSpam14 dataset contains the detailed labels of the tweets, \ie on which step a tweet was labeled during dataset construction). These training tweets are utilized to create the initial set of blacklisted domains, labeled clusters, trusted users, labeled tweets and spammy words. There are 48849 spam tweets and 22185 ham tweets. The remaining tweets on the first day and all the tweets of the remaining 14 days are used for testing purpose. Because not all tweets in HSpam14 are manually labeled, to ensure the accuracy of the evaluation in our experiments, we  manually label 300 randomly selected tweets from each time window to evaluate the performance of the system.\footnote{\scriptsize We have also evaluated the results using the manually labeled tweets in the HSpam14, similar results were obtained.} The performance is evaluated using the commonly used metric: Precision, Recall, and $F_1$.

Supervised spammer detection on Twitter such as~\cite{Lee:2010,Ferrara:2014} focus on spammer detection whereas our work is on tweet-level spam detection. Hence, these spammer detection systems can not be compared with our proposed method. Previous  tweet-level spam detection studies used off-the-shelf  classifiers namely, Na\"{\i}ve Bayes, Logistic Regression and  Random Forest classifier, in supervised settings. We have also used these methods  to compare  the performance with that of our proposed system. More specifically, tweet classification using logistic regression reported in this paper is similar to the work on information credibility~\cite{Castillo:2011} and also similar to the method reported in~\cite{Martinez:2013}. Most of the features described in the  information credibility paper except propagation related features are used in the \sdf as well. Propagation related features are not available in HSpam14 dataset so those features could not be used. Similarly, Na\"{\i}ve Bayes and  Random Forest classifiers are used in~\cite{Igor:2013,Martinez:2013,Wang:2015} for tweet-level spam detection. Random Forest classifier is found to be superior among all the  other methods~\cite{Igor:2013,Wang:2015}. In our experiments, we compare the results of \sdf with the classification results of using these classifiers in supervised setting.

\subsection{Result and Discussion}\label{sec:result}

\begin{table}[t]
\centering
\caption{Fraction of tweets detected by each detector}
\label{tab:eachStepTweet}
\begin{tabular}{l|l| r}
\hline
Step&   Detector & Coverage \\
\hline
   1& Blacklisted domain detection & 5.55\% \\
   2& Near-duplicates detection & 6.61\% \\
   3& Reliable ham tweets & 0.64\% \\
   4& Tweet classification & 87.20\% \\
\hline
\end{tabular}
\end{table}

\sdf has four detectors as shown in Figure~\ref{fig:systemOverview}. Table~\ref{tab:eachStepTweet} reports the percentage of tweets labeled by each detector. It shows that 5.55\% of the tweets are labeled by the blacklisted domain detector and 6.61\% of the tweets are labeled by the near-duplicate detector. Reliable ham tweet detector has very low coverage of 0.64\%. The low coverage is due to the fact that the HSpam14 dataset was collected based on popular hashtags, not on user basis~\cite{Sedhai:2015}. In other words, the dataset does not contain all tweets of any user. Because a trusted user should have at least 5 ham tweets, only a small set of users can be identified as trusted users.  Remaining 87.20\% of tweets are labeled by the last detector, tweet classifier. Next, we report the spam detection performance of \sdf with more focus on the tweet classifier detector.

\begin{figure*}[t]
 \centering
  \subfigure[Precision\label{sfig:precisionTweet}]
  {\includegraphics[width=1.9in]{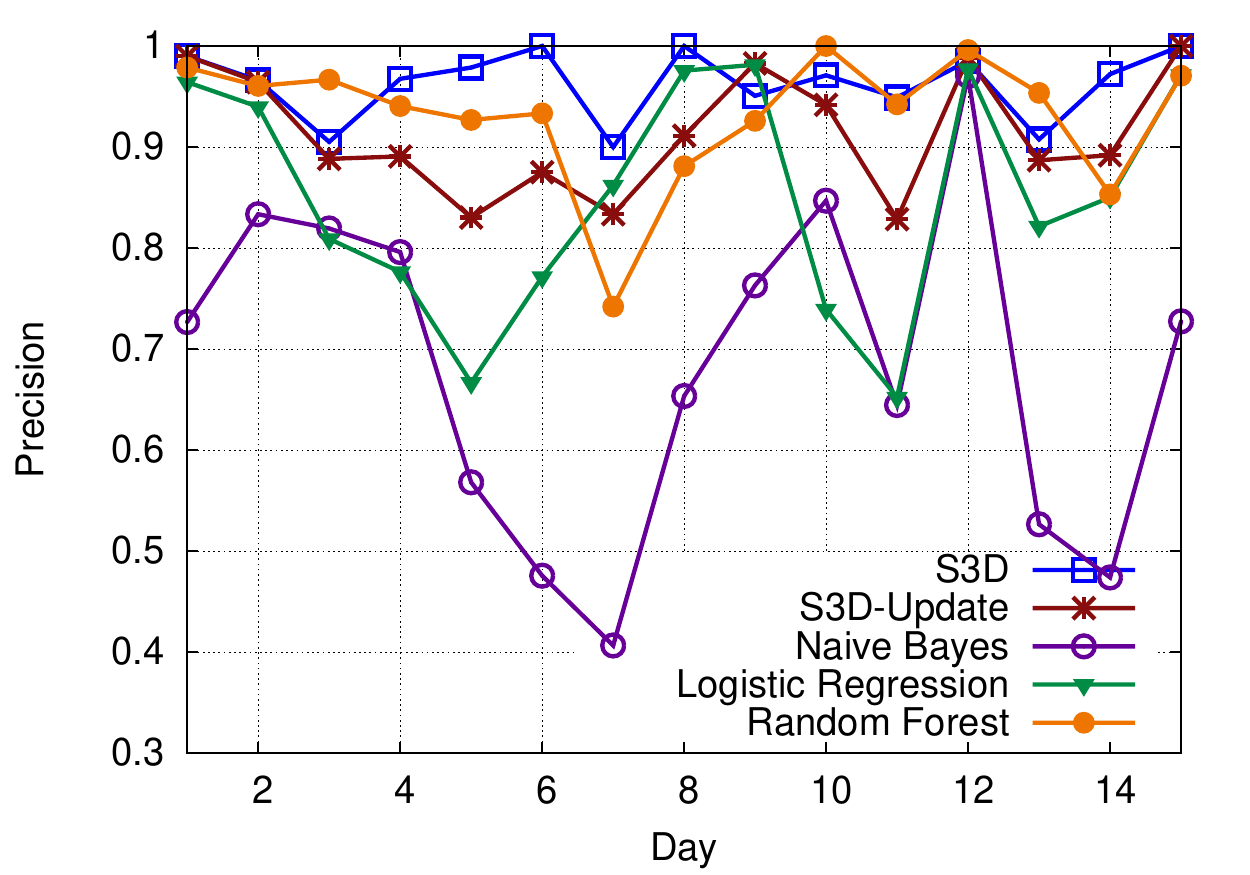}}
  \subfigure[Recall\label{sfig:recallTweet}]
  {\includegraphics[width=1.9in]{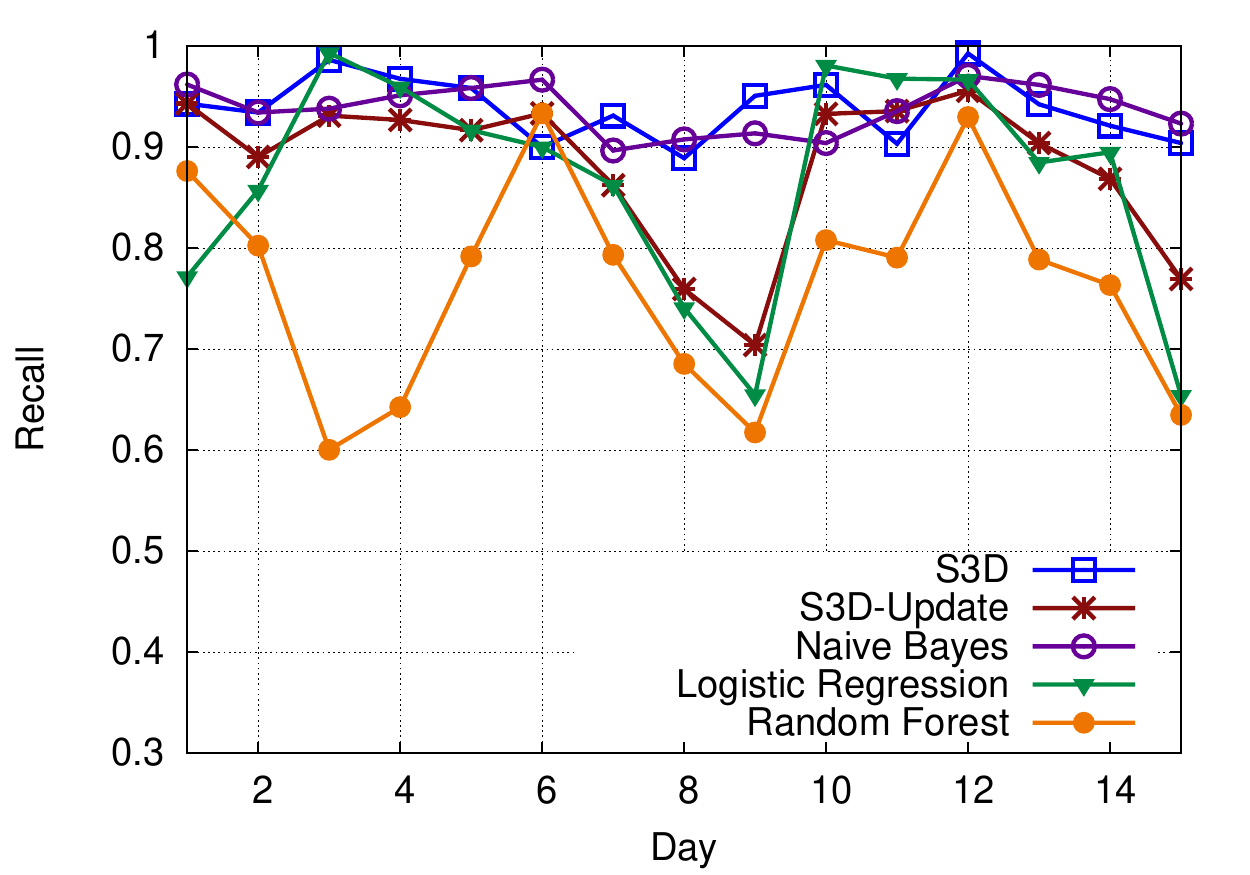}}
    \subfigure[$F_1$-score\label{sfig:f1-Tweet}]
  {\includegraphics[width=1.9in]{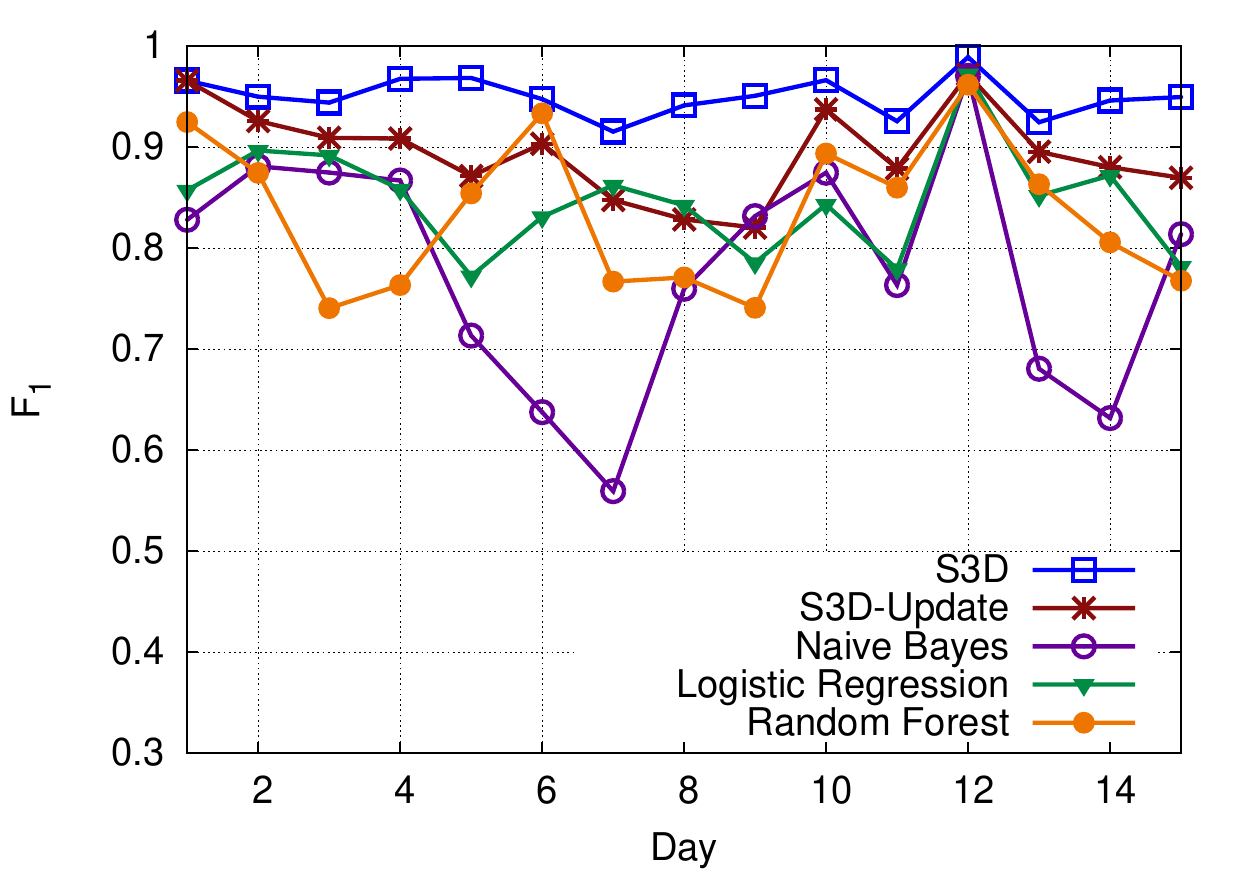}}
  \caption{Comparison of supervised and semi-supervised approach}
  \label{fig:evaluation}
\end{figure*}

\begin{figure}[t]
 \centering
  \subfigure[Domains, clusters, and users \label{fig:eachStepDCU}]
  {\includegraphics[width=1.9in]{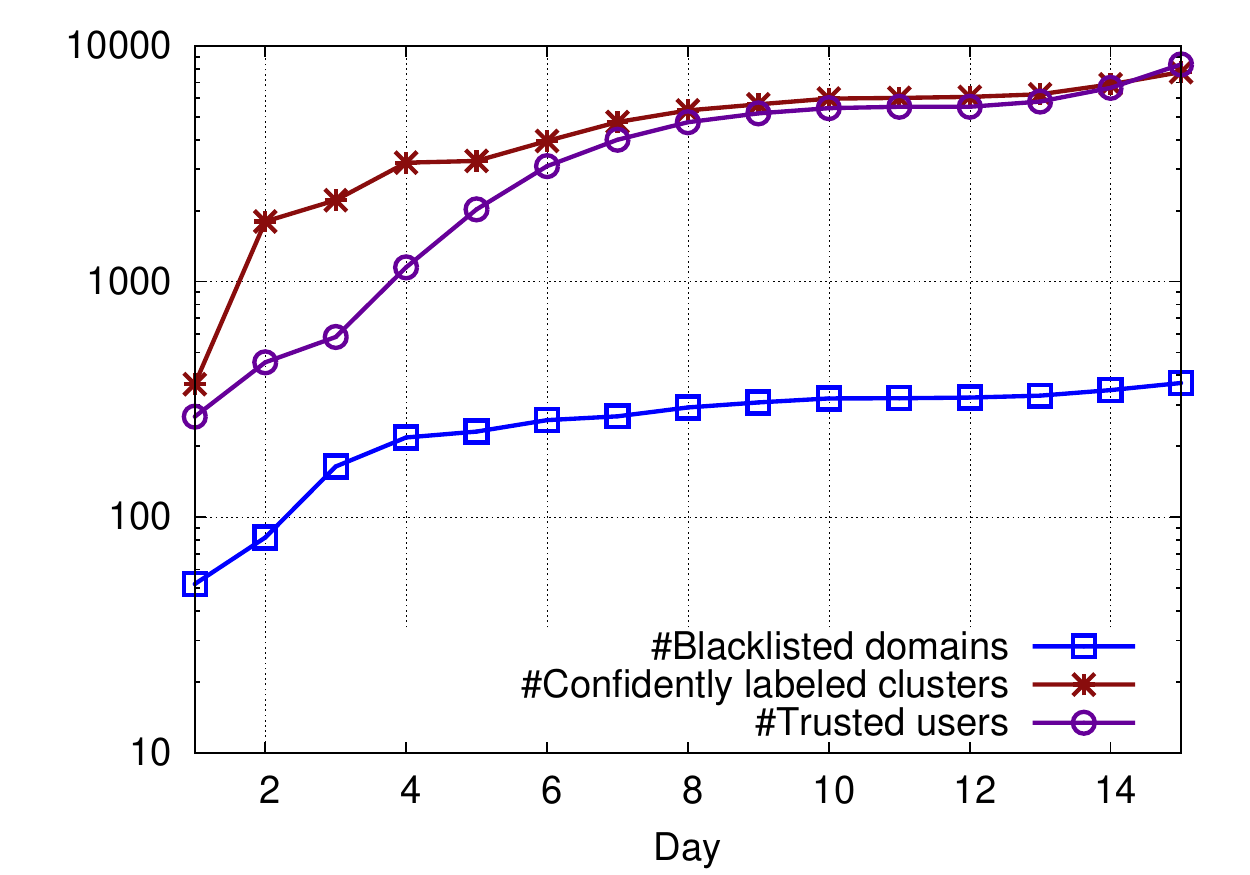}}
  \subfigure[Precision of the labels\label{fig:confidentClusterTweet}]
  {\includegraphics[width=1.9in]{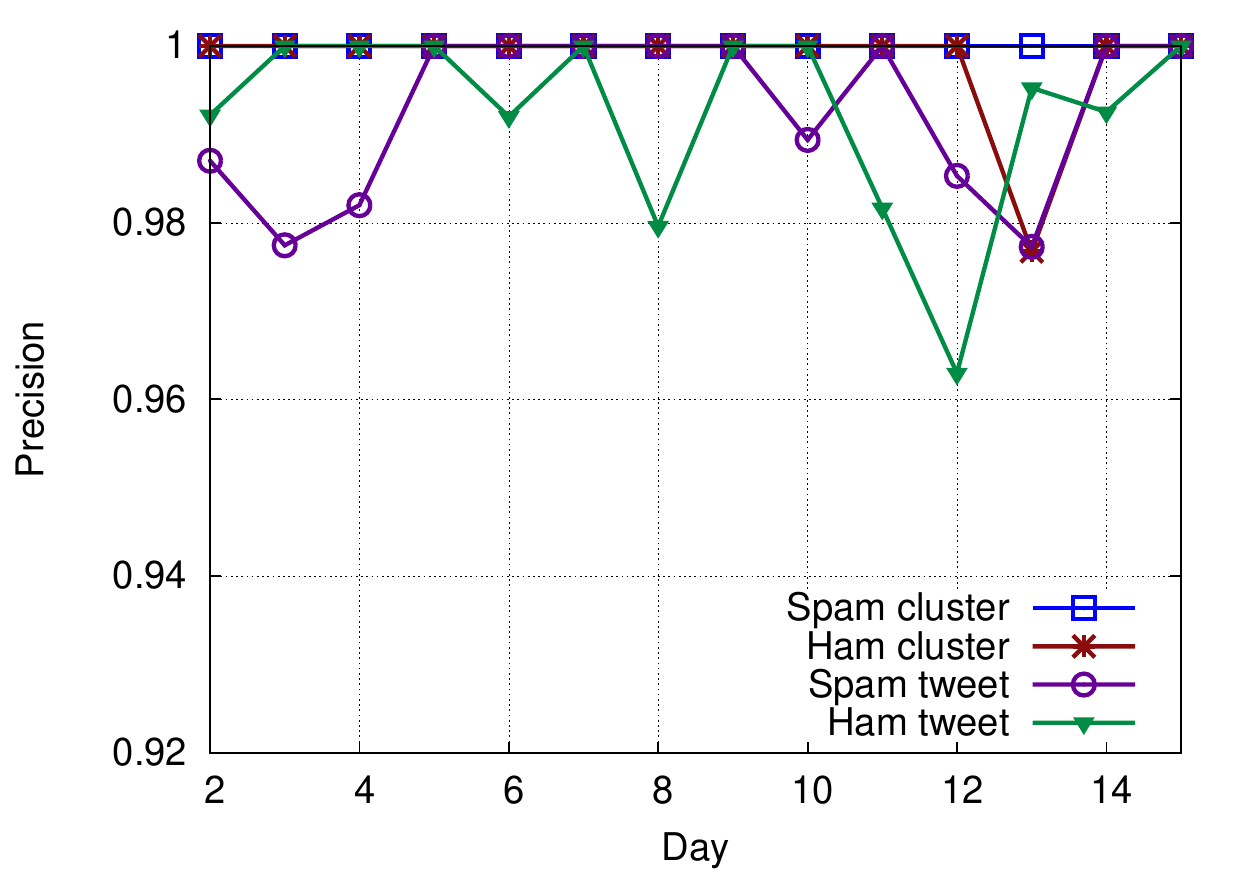}}
  \caption{Accumulated numbers of blacklisted domains, confident clusters and trusted users, and precision of confidently labeled clusters/tweets}
  \label{fig:evaluation2}
\end{figure}

We now report the performance of the following five methods for the spam tweet detection task.
\begin{itemize}
  \item \textbf{Na\"{\i}ve Bayes (NB)}: This method reports the prediction results of the Na\"{\i}ve Bayes classifier rely on the training data of the first day.
  \item \textbf{Logistic Regression (LR)}: This method reports the prediction results of the Logistic Regression classifier rely on the training data of the first day.
  \item \textbf{Random Forest (RF)}: This method reports the prediction results of the Random Forest classifier rely on the training data of the first day.
  \item \textbf{\sdf-Update}: The results of the \sdf framework without batch update. That is, the detectors in the framework fully rely on the training data of the first day, the same as the three classifiers above.
  \item \textbf{\sdf}: The results of the proposed \sdf framework, with model update after each time window.
\end{itemize}

Figure~\ref{fig:evaluation} plots the Precision, Recall and $F_1$ scores of the five methods. Observe \sdf achieves the best $F_1$ scores. The significant better $F_1$ scores against \sdf-Update over all days shows that semi-supervised approach is suitable for real-time spam detection in Twitter as it learns new spamming patterns continuously. Comparing \sdf with the  $F_1$  scores of NB, LR and RF shows that proposed method is superior to standard supervised methods. Observe $F_1$ scores of \sdf is consistent over time compare to other methods. It is observed that precision of RF is best  for 3 days but recall is the lowest of all. In contrast, NB has good recall at the expense of lower precision. The results show that proposed \sdf method is effective to capture spam tweets effectively.

The sudden rise of the $F_1$ scores on the 12$^{th}$ day is due to  a large number of relatively easy to detect spam tweets in that day. In HSpam14 dataset, the tweets were collected by using trending keywords of each day which leads to change in the distribution of words of each day.  The significant fluctuation of performance of \sdf may be due to the changing distribution of dataset in each time window. However, as \sdf continuously learns  new patterns and vocabulary,  its performance is found to be more consistent compare to other methods. More specifically, Figure~\ref{fig:eachStepDCU} plots the number of blacklisted domains, confidently labeled near-duplicate clusters and trusted users. It shows that \sdf keeps on utilizing new knowledge obtained from earlier labeled tweets and clusters to improve the capability of spam tweets detection. Furthermore, we have also used top 10,000 frequent uni-/bi- and tri-grams  computed at the end of each time window to update the model to deal with vocabulary change (see Table~\ref{tab:features}).

\begin{table}
\centering
\caption{Top-15 most effective vocabulary features}
\label{tab:TFImportance}
\begin{tabular}{l|p{2.5in}}
\hline
Tweet&   follow; follow back; back; ipad ipadgames;  please follow;  please;
   followers; retweet; tfbjp; teamfollowback;   ipad;  follow me;    collected;
   followback;   gameinsight\\ \hline
Cluster&
follow; love; back; followers; follow me; follow back; retweet; openfollow; teamfollowback;
want; someone; win; please; 500aday; gain\\

\hline
\end{tabular}
\end{table}

In \sdf, we identify the confidently labeled tweets and the confidently labeled near-duplicate clusters. The confidently labeled tweets and clusters are utilized to learn new models for the detectors. The quality of these confidently labeled tweets and clusters are therefore crucial for the performance of \sdf. Here, we evaluate the quality of these tweets and clusters, plotted in Figure~\ref{fig:confidentClusterTweet}.  The confident clusters are evaluated by manually labeled 47 randomly selected clusters on each day, which is the smallest number of confidently labeled clusters produced over the 15 days.
 Figure~\ref{fig:confidentClusterTweet} shows that the precision of confident clusters is almost perfect for both spam and ham clusters.  The figure also shows that the precision of confidently labeled spam and ham tweets are consistently above 95\%. Adding such clusters and tweets in the training process makes  \sdf capable of capturing emerging spamming activities as well as new vocabulary.

\begin{table*}
\centering
\scriptsize
\caption{Features used to represent tweets and clusters for classification; \textbf{FoT} means \textit{\textbf{F}raction \textbf{o}f \textbf{T}weets} and \textbf{FoU} means \textit{\textbf{F}raction \textbf{o}f \textbf{U}sers}.   The top-15 most effective features for tweet
classification and cluster classification based on the Gini impurity score are indicated by the numbers “($x$)” following the features ($1\leq x \leq 15$).
}
\label{tab:features}
\begin{tabular}{l |p{2.6in} |p{2.8in}}
	\hline
	Type & Features for tweet representation  &  Features for cluster representation \\
	\hline
\textbf{Hashtag}& Contains hashtag  &      FoT having hashtag (3)\\
  & Contains more than 2  hashtags (1)  &   FoT having more than 2 hashtags (5)\\
  & Contains spammy hashtag (2)   &  Hashtag per tweet (12)\\

  & Contains categorical hashtag (7)  &        FoT having spammy hashtag (14)  \\
  & Contains capitalized hashtag  (8)& FoT having categorical word as hashtag \\
  & & FoT having capitalized hashtag (4)\\

\hline
\textbf{Content}    & Fraction of words that are spammy &                 FoT having spammy words  \\
  & Contains question mark &                            FoT having question mark \\
  & Contains money sign &                               FoT having exclamation mark (10)\\
  & Contains exclamation sign (13) &                         FoT having money sign  \\
  & Contains positive emoticons &                       FoT having positive emoticons (7)  \\
  & Contains negative emoticons &                       FoT having negative emoticons \\
  & Contains positive words &                           FoT having positive words \\
  & Contains negative words  &                          FoT having negative words (8) \\
  & Fraction of uppercase characters (9)  &                Fraction of capitalized tweets (2) \\
  & Contains URL (11)&                                      FoT that are retweet (6)\\
  & Is retweet &                                        FoT having URL \\
  & Contains mention (14) &                                 Mentions per tweet (15)\\
  & Contains first person pronoun  &                    FoT tweets having first person pronoun \\
  & Contains second person pronoun &                    FoT tweets having  second person pronoun \\
  & Contains third person pronoun &                     FoT tweets having third person pronoun \\
  & Normalized length of the tweet in word &            Median tweet length in word/(max tweet length) \\
  & Normalized length of the tweet in character (10) &       Median tweet length in characters /140 \\
  & Contains top 10,000 uni-/bi-/tri-gram of last time window  &      Contains top 10,000 uni-/bi-/tri-gram of the last time window\\
  & Day of the week in which the tweet is posted  & Ratio of spam tweets in the cluster (1)\\
\hline	
\textbf{User}  & Has less than 5 percentile followers (3)  &         FoU having less then 5 percentile followers \\ %
 & Has less than 5 percentile followees (6) &         FoU having less then 5 percentile followees  \\
 & Has more than 50 percentile total tweets &      FoU having more than 50 percentile total tweets \\
 & Percentile followers of the user (15) &           Median  percentile of followers of users \\
 & Percentile followees of the user &              Median  percentile of followees of users \\
 & Percentile total tweet count of the user &      Median percentile of total tweets of the users (11)\\
 & User profile contains description (5) &             FoU having description in profile \\
 & User profile description contains spammy words &FoU having spammy words in description\\
 & User profile has url  &                         FoU having URL in profile \\
 & User profile has location info  &               FoU having location info in profile \\
 & User profile has time-zone info &              FoU having timezone info in profile \\
 & Followers-followees ratio (4) &                     FoU having followers greater than followee \\
 & User's normalized age (12) &                         Median normalized age of users \\
 & & Fraction of post by the dominating user (9)\\
 & & Percentile followers of the user tweeting the most \\
 & & Percentile followees of the user tweeting the most\\
 & & Percentile total tweets of the user tweeting the most \\
 & & Tweets per user \\
 & & Standard deviation of normalized age of users (13) \\
 & & Percentile  followers of the most followed user \\
 & & Percentile  followees of the most followed user\\
 & & Percentile total tweets of the most followed user \\
\hline
\textbf{Domain} & URL from top 100 domains  & FoT having URL from top 100 domains  \\
  & URL from top 1000 domains & FoT having URL from top 1000 domains \\
  & URL from top 10000 domains & FoT having URL from top 10000 domains \\
	\hline
\end{tabular}
\end{table*}

\subsection{Feature Analysis}

There are four types of features used to represent tweet and cluster for classification (see  Table \ref{tab:features}). In our experiments, we observe that normalization of features gives better performance than without normalization. Because users' followers, followees and total tweets exhibit power law distributions. The features derived from these values are normalized based on percentile. Features such as length of a tweet in characters and words show normal distribution which are normalized by the maximum value. Based on the Gini impurity score, we identify the top-15 most effective features for tweet classification and cluster classification respectively. These features are highlighted in Table~\ref{tab:features} in ``($x$)'' format, where $x$ is the top ranking position.

It is observed that 10 out of the top-15  most effective features are vocabulary-based features for tweet classification whereas in the case of cluster classification only 3 out of the top-15  features are vocabulary-based features.  Meta-data of a tweet contains  information only about the single tweet which is comparatively less informative. In contrast, a cluster contains a number of tweets, hence meta-data based features represent the collective information of tweets in the cluster and are comparatively more informative. For example, if there is a tweet from a user whose account creation date is known and  has very few followers and followees, it is hard to determine that tweet posted by this users is ham or spam. In contrast, if there is a  group of users whose accounts are created around the same time and all having very few number of followers and followees, and posting near-duplicate tweets, then the tweets in this cluster are likely to be spam.

Table~\ref{tab:TFImportance} lists the top-15 vocabulary-based features (uni-gram, bi-gram and tri-gram). Only uni-gram and bi-gram vocabulary appear in the top ranked list. One possible reason for tri-gram features not in the list may be due to the sparsity of the tri-gram vocabulary in the dataset. It is interesting to note that most of the top words based on Gini-impurity score are the same as the list of hashtags having the highest \textit{spammy-index} reported in~\cite{Sedhai:2015}.

\section{Conclusion}\label{sec:conclusion}
In this paper, we propose a semi-supervised spam detection framework, named \sdf. \sdf utilizes four light-weight detectors to detect spam tweets on real-time basis and update the models periodically in batch mode. The experiment results demonstrate the effectiveness of semi-supervised approach in our spam detection framework. In our experiment, we found that confidently labeled clusters and tweets make the system effective in capturing new spamming patterns. Due to the limited user information in the dataset, we have used the simple technique to deal with user-level spam detection. However, we argue that the user-level spam detection can be incorporated into \sdf, which is part of our future work.

{\small
\bibliographystyle{abbrv}
\bibliography{S3D-TCSS}

\begin{thebibliography}{10}

\bibitem{Benevenuto:2010}
F.~Benevenuto, G.~Magno, T.~Rodrigues, and V.~Almeida.
\newblock Detecting spammers on twitter.
\newblock In {\em {CEAS}}, 2010.

\bibitem{Broder:1997}
A.~Broder.
\newblock On the resemblance and containment of documents.
\newblock In {\em Proc. Compression and Complexity of Sequences}, pages 21--29,
  1997.

\bibitem{Castillo:2011}
C.~Castillo, M.~Mendoza, and B.~Poblete.
\newblock Information credibility on twitter.
\newblock In {\em {WWW}}, pages 675--684, 2011.

\bibitem{Chen:2015}
C.~Chen, J.~Zhang, Y.~Xie, Y.~Xiang, W.~Zhou, M.~M. Hassan, A.~AlElaiwi, and
  M.~Alrubaian.
\newblock A performance evaluation of machine learning-based streaming spam
  tweets detection.
\newblock {\em IEEE Transactions on Computational Social Systems}, 2(3):65--76,
  2015.

\bibitem{Chhabra:2011}
S.~Chhabra, A.~Aggarwal, F.~Benevenuto, and P.~Kumaraguru.
\newblock Phi.sh/\$ocial: The phishing landscape through short urls.
\newblock In {\em {CEAS}}, pages 92--101, 2011.

\bibitem{Chu:2010}
Z.~Chu, S.~Gianvecchio, H.~Wang, and S.~Jajodia.
\newblock Who is tweeting on twitter: Human, bot, or cyborg?
\newblock In {\em Proc. Annual Computer Security Applications Conf.}, pages
  21--30, 2010.

\bibitem{Ferrara:2014}
E.~Ferrara, O.~Varol, C.~Davis, F.~Menczer, and A.~Flammini.
\newblock The rise of social bots, 2014.
\newblock arXiv:1407.5225.

\bibitem{Gao:2010}
H.~Gao, J.~Hu, C.~Wilson, Z.~Li, Y.~Chen, and B.~Y. Zhao.
\newblock Detecting and characterizing social spam campaigns.
\newblock In {\em {IMC}}, pages 35--47, 2010.

\bibitem{Ghosh:2012}
S.~Ghosh, B.~Viswanath, F.~Kooti, N.~K. Sharma, G.~Korlam, F.~Benevenuto,
  N.~Ganguly, and K.~P. Gummadi.
\newblock Understanding and combating link farming in the twitter social
  network.
\newblock In {\em {WWW}}, pages 61--70, 2012.

\bibitem{Grier:2010}
C.~Grier, K.~Thomas, V.~Paxson, and M.~Zhang.
\newblock @spam: The underground on 140 characters or less.
\newblock In {\em Proc. ACM Conf. on Computer and Communications Security},
  pages 27--37, 2010.

\bibitem{Hu:2014b}
X.~Hu, J.~Tang, and H.~Liu.
\newblock Online social spammer detection.
\newblock In {\em {AAAI}}, pages 59--65, 2014.

\bibitem{Hu:2013}
X.~Hu, J.~Tang, Y.~Zhang, and H.~Liu.
\newblock Social spammer detection in microblogging.
\newblock In {\em {IJCAI}}, pages 2633--2639, 2013.

\bibitem{Lee:2010}
K.~Lee, J.~Caverlee, and S.~Webb.
\newblock Uncovering social spammers: Social honeypots + machine learning.
\newblock In {\em {SIGIR}}, pages 435--442, 2010.

\bibitem{Martinez:2013}
J.~Martinez-Romo and L.~Araujo.
\newblock Detecting malicious tweets in trending topics using a statistical
  analysis of language.
\newblock {\em Expert Systems with Applications}, 40(8):2992--3000, 2013.

\bibitem{Igor:2013}
I.~Santos, I.~Mi{\~{n}}ambres{-}Marcos, C.~Laorden,
  P.~Gal{\'{a}}n{-}Garc{\'{\i}}a, A.~Santamar{\'{\i}}a{-}Ibirika, and P.~G.
  Bringas.
\newblock Twitter content-based spam filtering.
\newblock In {\em Joint Conf. SOCO'13-CISIS'13-ICEUTE'13}, pages 449--458,
  2013.

\bibitem{Santos:2011}
I.~Santos, J.~Nieves, and P.~G. Bringas.
\newblock {\em Semi-supervised Learning for Unknown Malware Detection}, pages
  415--422.
\newblock 2011.

\bibitem{Sedhai:2015}
S.~Sedhai and A.~Sun.
\newblock Hspam14: A collection of 14 million tweets for hashtag-oriented spam
  research.
\newblock In {\em {SIGIR}}, pages 223--232, 2015.

\bibitem{Tan:2013}
E.~Tan, L.~Guo, S.~Chen, X.~Zhang, and Y.~Zhao.
\newblock Unik: Unsupervised social network spam detection.
\newblock In {\em {CIKM}}, pages 479--488, 2013.

\bibitem{Thomas:2011}
K.~Thomas, C.~Grier, J.~Ma, V.~Paxson, and D.~Song.
\newblock Design and evaluation of a real-time url spam filtering service.
\newblock In {\em IEEE Symposium on Security and Privacy}, pages 447--462,
  2011.

\bibitem{Wang:2015}
B.~Wang, A.~Zubiaga, M.~Liakata, and R.~Procter.
\newblock Making the most of tweet-inherent features for social spam detection
  on twitter.
\newblock {\em CoRR}, abs/1503.07405, 2015.

\bibitem{Wu:2009}
Y.~S. Wu, S.~Bagchi, N.~Singh, and R.~Wita.
\newblock Spam detection in voice-over-ip calls through semi-supervised
  clustering.
\newblock In {\em 2009 IEEE/IFIP International Conference on Dependable Systems
  Networks}, pages 307--316, 2009.

\bibitem{Yang:2011}
C.~Yang, R.~Harkreader, and G.~Gu.
\newblock Die free or live hard? empirical evaluation and new design for
  fighting evolving twitter spammers.
\newblock In {\em Recent Advances in Intrusion Detection}, volume 6961 of {\em
  Lecture Notes in Computer Science}, pages 318--337. 2011.

\end{thebibliography}
}

\end{document}